\documentstyle[12pt]{article}

\def\Tr{{\rm Tr}}

\begin{document}
\begin{center}
{\LARGE{\bf  Quantum Evolution and Space-time Structure}}
\vskip 0.5cm
quant-ph/9512004
\vskip 1cm

{\large {\bf George Svetlichny}}

\vskip 0.5 cm

Departamento de Matem\'atica \\
Pontif\'{\i}cia Universidade Cat\'olica \\
Rua Marqu\^es de S\~ao Vicente 225 \\
22453 G\'avea, Rio de Janeiro, RJ, Brazil

\end{center}

\begin{abstract}
The hilbert-space structure of quantum mechanics is related to the
causal structure of space-time. The usual measurement hypotheses
apparently preclude nonlinear or stochastic quantum evolution. By
admitting a difference in the calculus of joint probabilities of events
in space-time according to whether the separation is space-like or
time-like, a relativistic nonlinear or stochastic quantum theory may be
possible.
\end{abstract}

\section{Nonlinear and stochastic quantum evolution}
The motivation for considering nonlinear or stochastic quantum evolution
is varied: fundamental speculation, presence of gravity, string theory,
representations of current algebras, etc.  On the other hand it is
becoming progressively evident that nonlinear quantum mechanics (and
possibly stochastic also) is a radical departure from conventional
theory.  This has already become apparent more than a decade ago.
According to Bugajski (\cite{Bug} and earlier references therein) such
theories are classical theories with constraints, situated somewhere
between classical and quantum mechanics.  Haag and Bannier (\cite{HaBa})
pointed out that in Mielnik's nonlinear quantum mechanics (\cite{Mie})
one can distinguish between two different convex combinations of pure
states that lead to the same density matrix. This makes the state-space
a simplex, just as in classical theories, and, as it became apparent
later, allows for superluminal signals. As has been pointed out by N.
Gisin (\cite{Gis1,Gis2}) and G. Svetlichny (\cite{Sve1}) nonlinearity
allows us to use EPR-type correlations and the instantaneous nature of
state-vector collapse to send a signal across a space-like interval.
Polchinski (\cite{Pol}) argues that in  Weinberg's nonlinear theory
(\cite{Wei1,Wei2}) one can either communicate between separate branches
of an Everett multiple-world universe or physical systems can react to
the content of the experimenter's mind.  Further peculiarities are that
even for non-interacting systems, higher particle-number equations are
not uniquely determined by one-particle equations (\cite{GoSv}),
and that such ambiguities become important for
particles with internal symmetries. In fact there are non-trivial
obstruction to lifting symmetries from \(N\)-particles to
\(N+1\)-particles (\cite{Sve2}).

\section{Relativity constraints and problems}

The presence of superluminal signals in nonlinear theories was the first
indication that nonlinearity and relativity are in conflict. In fact the
presence of such signals {\em per se\/} already contradicts relativity.
To make this clear, consider a superluminal signaling device set up
according to the state-collapse mechanism and that is to operate between
two distant locations in the rest reference frame of two observers at
relative rest. According to the mechanism explained in the cited
articles, if at \(t=0\) the first observer changes over from measuring
one observable to a suitable other, then the second observer, given a
nonlinear time evolution, will, after a negligible time interval, detect
a change in the expected value of the observable he is measuring and
consequently receive a signal. We can say that for the second observer
the {\em onset\/} of the signal is at \(t=\epsilon > 0\) for some small
\(\epsilon\). Onset is a physical event and so all observers ought to
agree where in space-time it occurred. Consider how the same situation
is seen in a reference frame of a moving observer. He would see a
different initial state, find that the time-evolution is given by a
possibly different nonlinear equation, and if special relativity holds,
that collapse occurs in a different plane of simultaneity. The argument
that leads to superluminal signals is sufficiently general that the
moving observer will also expect these to exist, but now in relation to
his plane of simultaneity, and so he would expect the onset of the
signal along the second observer's world-line to to be significantly
different from what was determined before. Since onset is an
uncontestable physical fact,  this is a contradiction.

Relativity constraints on quantum evolution is something that has not
yet been fully explored. The problem arises with the measurement
process. Consider a measurement with space-like separated instrumental
events such as a correlation measurement upon a two particle system of
the EPR type. In one frame the measurements on the two particles are
simultaneous and so can be considered as just parts of a single
measurement, while in another frame the two measurements are successive
with intervening time evolution. These two description must be
equivalent and produce the same observable results. Thus relativity
imposes constraints that relate the measurement process to the
evolution. These constraints pose obvious difficulties for stochastic
evolution, for in the frame where there is a single measurement the
outcomes can be calculated from the measurement process algorithm
applied to the state just prior to the measurement. In the other frame
there is an intervening dissipative evolution, a dissipation not present
in the first frame. It is questionable that one can maintain an
equivalence of the two descriptions. That there are also difficulties
for nonlinear evolution is not as apparent but they do exist and we
shall refer to them later.

Another, but related, constraint comes about in considering a
measurement process in a limited space-time region and two observers in
relative motion at  space-like  separation from the measurement region
such that for one observer the measurement has already taken place while
for the other it has not. One observer would subject his
state-description to a collapse while the other would not.  These
different descriptions must not have local observable effects and this
is a constraint on the theory.

Another hint of these difficulties can be seen by considering the
following commutator in the lie algebra of the Poincar\'e group.
\[[L_{0j},P_j] = P_0, \]
that is, the commutator of a boost generator and the collinear momentum
is the energy. The moral is that one cannot impose on the time
evolution, properties that one would not impose on neither space
translation nor boosts.

One thus comes to the realization that {\em for a relativistic nonlinear
or stochastic quantum theory to be viable the measurement process must
be modified\/}.  Once this is realized one must be aware that it is very
easy to make certain types of trivial modifications. Let \(T:H \to H\)
be a nonlinear invertible norm-preserving transformation of a Hilbert
space \(H\).  Let \(U(t)\) be a unitary quantum evolution operator and
\(P\) a spectral projector of an observable. One has an obvious
equivalence between the evolution and measurement processes as described
by  the two sides of the following diagram:
\[\begin{array}{c} \Psi \mapsto U(t)\Psi \\ \Psi \mapsto P\Psi
\end{array} \Leftrightarrow \begin{array}{c}
T\Psi \mapsto TU(t) \Psi\\
T\Psi \mapsto TP\Psi\end{array}
\]
What one has done on the right-hand side is introduced curvilinear
coordinates in Hilbert space but left physics alone.  There are two ways
of avoiding triviality. One would be to leave part of the formalism
unmodified, such as in those proposals that modify the evolution but
maintain the usual measurement process. The difficulty of this is that
one runs the risk of contradiction. The other way is to deal only with
invariant objects such as  joint probability distributions of events in
space-time. This is notoriously difficult but is the only way to achieve
true insight into the problem.

\section{Joint probabilities in quantum mechanics}

Consider successive measurements with finite spectrum
operators,
\[
A = \sum_i \lambda_i P_i,\quad \quad\quad
B = \sum_j \mu_j Q_j,
\]
performed on a (possibly mixed) heisenberg state represented by the
density matrix   \(\rho_0\).
The joint probability of seeing outcomes \((i,j)\) for the two
measurements is
\[P(i,j)=\Tr (Q_jP_i\rho_0P_iQ_j).\]
and the conditional probabilities are:
\begin{eqnarray}
P(j|i) &=& \frac{\Tr (Q_jP_i\rho_0P_iQ_j) }{\Tr (P_i\rho_0)} \\
P(i|j) &=& \frac{\Tr (Q_jP_i\rho_0P_iQ_j) }{\sum_k\Tr
(Q_kP_i\rho_0P_iQ_k)}
\end{eqnarray}
Conditional probabilities are important in that they often correspond to
what is measured in the laboratory. Very often in practice one does not
execute the observation procedure only in the instances that the
preparation procedure is deemed successful. What does take place is that
one performs a long experimental run and only {\em a posteriori\/}
analyses those instances in which the preparation was deemed successful.
This is most apparent for instance in high-energy physics. A simple
model for what happens in practice would be to consider that there is
some ``gross" preparation procedure and two observation procedures. A
long experimental run is executed and only the cases in which a
particular outcome in one of the observations is realized are considered
to be the cases in which the desired state of affairs has been created
and for which the outcomes corresponding to the other observation
procedure are then subsequently analyzed. Data for which some other
outcome of the first observation is obtained are simply ignored. The
procedure describe above can be called an {\em indirect\/} preparation
procedure. The normal attitude concerning it is that the compound
procedure ``execute a preparation procedure then execute an observation
procedure and consider the operation successful if such and such outcome
obtains" is a procedure just as legitimate for creating a state of
affairs as any other. One collects data even if the indicated outcome,
which we shall call the {\em conditioning\/} outcome,
did not occur, merely for technological reasons, it would just be too
difficult or impossible to set up the experiment in another way. Since
by assumption the separate execution of the experiment in the long run
do not interfere with each other, the fact that the instances of the
desired state of affairs are imbedded in a larger set along with states
of affairs of no interest is innocuous as mere data analysis weeds them
out. Consider now the two observations introduced above in this light and
consider one of them as the conditioning observation for an indirect
state preparation.
Now it is usual to consider the conditioning observations as taking
place before the conditioned observation, what can be called {\em
pre-conditioning\/} but since one performs the data analysis after all
the data has been collected one could perform, {\em post-conditioning\/},
that is,  conditioning on future events.  It is instructive to contrast
the two:

{\em Pre-conditioning:}
\[P(j|i) = \Tr (Q_j \rho_i)\]
 where \(\rho_i = P_i\rho_0P_i /
\Tr(P_i\rho_0)\).
\begin{itemize}
\item The new density matrix \(\rho_i\) depends only on \(P_i\) and not
on the other compatible spectral projectors \(P_k\), \(k \neq i\).
\item Given \(\rho_i\), \(P(j|i)\) depends only on \(Q_j\) and not
on the other compatible spectral projectors \(Q_k\), \(k \neq i\).
\end{itemize}
{\obeylines {\em Post-conditioning:}}

Unless \([A,B]=0\), \(P(i|j)\) depends not only on \(\rho_0\), \(P_i\),
and \(Q_j\) but also on the other projectors in the two spectral
decompositions.
\begin{itemize}
\item The ``state of affairs" created by post-conditioning on outcome
\(j\) depends on the outcome's ``context", the other compatible
projections \(Q_k\), \(k \neq i\). {\em Contextual Conditioning}
\item The above ``state of affairs" breaks the equivalence class of
experimental outcomes where two such are equivalent if they correspond
to the same spectral projector.
\end{itemize}

One sees that for commuting observables, post-conditioning behaves
exactly the same as pre-conditions. This means also that {\em
space-like\/} conditioning behaves the same as time-like
pre-conditioning. This last statement is a characteristic of quantum
mechanics and may in fact be a determining condition in a relativistic
theory.  One can show (\cite{Sve3}) that lorentz covariance imposes
constraints on joint probabilities of events in space-time:
Let \(I\) and \(J\) be two space-like separated instruments with outcomes
\(\{a_1,\dots,a_n\}\) and \(\{b_1,\dots,b_m\}\) then,
\begin{eqnarray*}P_{i,j}^{I \wedge J}({\cal W}) &=&
P_j^J(\pi^I_i {\cal W})P_i^I({\cal W}) \\
\pi^{I \wedge J}_{i,j} &=& \pi^J_j \pi^I_i
\end{eqnarray*}
where \(P\) is probability \({\cal W}\) is a preparation procedure and
\(\pi\) is the conditioning operator for indirect preparation.
These constraints imposed in their {\em non-contextual} form on
(adequately defined) {\em compatible} instruments lead in several
axiomatic schemes (\cite{Poo1,Poo2,Guz1,Guz2,Gis3})
to a
hilbert-space model for physical propositions. From here one has
arguments that lead to linearity of evolution (\cite{Gio1,Gio2,Jor}).

The moral here seems to be that there is a relation (independently
postulated by N. Gisin and G. Svetlichny) between space-time
structure (relativistic causality in particular) and the hilbert-space
model of quantum mechanics. The fact that one must impose
the relativistic constraints on all pairs of compatible instruments and
not only on the space-like separated ones has two implications. The
first is that the identical behaviour of space-like conditioning and
time-like pre-conditioning may be, along with lorentz covariance, a
determining condition for hilbert-space quantum mechanics. Thus one may
conjecture that any relativistic theory with non-contextual
conditioning for future measurements (whether time-like or space-like)
must be a Hilbert-space theory (with possible superselection sectors)
with linearly implemented (probably deterministic) time evolution.
The second implication is that for a relativistic nonlinear quantum
mechanics to be possible, one probably has to introduce a discontinuity
in the conditioning behavior for indirect preparations across the
future-light cone and allow space-like conditioning to behave
differently from future time-like.

Another possibility for a nonlinear theory would be to modify the
measurement process to be contextual (as happens for post-conditioning)
but still maintain that space-like and future time-like conditioning
follow the same rules. Unfortunately we have no general results
concerning this possibility though some preliminary results suggest that
such theories face the same difficulties as the nonlinear
non-contextual ones.

\section{Possibilities for nonlinear relativistic quantum mechanics}

{}From the discussion of the previous section one can conjecture that a
nonlinear relativistic quantum mechanics can be achieved if space-like
and future time-like conditioning behave differently. Since space-like
cannot be changed to time-like by a lorentz transformation, the proposal
does not conflict with relativity, at least not superficially. The
proposal avoids superluminal signals since these would only be related
to space-like conditioning which would have to obey the constraints of
the previous section which already preclude such signals (\cite{Sve3}).
What must then be modified is the future time-like conditioning. To get
some idea of such possible modification consider a free neutral scalar
relativistic quantum field.  For each limited space-time region \({\cal
O}\) let \({\cal A}({\cal O})\) be the algebra of observables associated
to \(\cal O\). Consider now a set of limited space-time regions \({\cal
O}_1,\dots{\cal O}_n\) which are so disposed that for any two, either all
points of one are space-like in relation to all points of the other, or
they are time-like.  Assume the regions are numbered so that whenever
one is in the time-like future of another, then the first one has a
smaller index. Let \(P_i \in {\cal A}({\cal O}_i)\) be orthogonal
projections that correspond to  outcomes of measurements made in the
corresponding regions.  Let \(\Psi\)  represent a heisenberg state in
some reference frame and prior to all measurements.  According to the
usual rules, the probability to obtain all the outcomes represented by the
projections is: \[||P_1P_2\cdots P_n\Psi||^2.\]
A modification of the sort we are proposing would be, for instance,
to replace in this
formula \(P_i\) by \(P_iB_i\) whenever there is a region \({\cal O}_j\)
that is time-like past to the given one. This effectively differentiates
between space-like and time-like conditioning. For this to be consistent
and relativistic the presumably nonlinear operators \(B_i\) would
have to satisfy certain constraints. If we can associate to a space-time
region \({\cal O}\) a possibly nonlinear operator \(B_{\cal O}\) such
that
\begin{enumerate}
\item Operators assigned to space-like separated regions commute and
the operator assigned to a region commutes with all projectors associated
to a space-like separated region.
\item If \({\cal O} \subset {\cal O}'\)
and \(P \in {\cal A}({\cal O})\) is a projector then \(PB_{{\cal O}'} =
PB_{\cal O}\)
\item If \(U(g)\) is a unitary operator representing the element \(g\)
of the Poincar\'e group then \(B_{g{\cal O}} = U(g)^*B_{\cal O}U(g)\)
\end{enumerate}
then the above prescription would already constitute a nonlinear
relativistic quantum theory. One still does not know how to compute
joint probabilities for events in regions that are neither space-like nor
time-like to each other but the case at hand would certainly have to
addressed and would constitute a first step. It is not yet know if an
association \({\cal O} \mapsto B_{\cal O}\) satisfying these constraints
exists. Even if it does not, the path toward a relativistic non-linear
quantum mechanics is now sufficiently clear that one may feel that such
a mechanics may after all be possible in spite of the weighty arguments
brought forth against it up to now.

\section*{Acknowledgment}
The author thanks the Arnold Sommerfeld Institute for its kind invitation
to the symposium and for financial support. This research was financed
by the Minist\'erio de Ci\^encia e Tecnologia (MCT) and by the Conselho
Nacional de Desenvolvimento Tecnol\'ogico e Cient\'{\i}fico (CNPq),
agencies  of the Brazilian government.

\end{document}